\documentclass[aps,prl,twocolumn,superscriptaddress,showpacs]{revtex4}
\usepackage{epsfig}
\usepackage{amsmath}
\usepackage{times}

\begin{document}

\title{An efficient approach of controlling traffic congestion in scale-free networks}

\author{Zonghua Liu}
\affiliation{Institute of Theoretical Physics and Department of
Physics, East China Normal University, Shanghai, 200062, China}
\author{Weichuan Ma}
\affiliation{Department of Physics, Hubei University, Wuhan,
430062, China}
\author{Huan Zhang}
\affiliation{Institute of Theoretical Physics and Department of
Physics, East China Normal University, Shanghai, 200062, China}
\author{Yin Sun}
\affiliation{Institute of Theoretical Physics and Department of
Physics, East China Normal University, Shanghai, 200062, China}
\author{P. M. Hui}
\affiliation{Department of Physics, The Chinese University of Hong
Kong,\\ Shatin, New Territories, Hong Kong}

\date{\today}

\begin{abstract}
We propose and study a model of traffic in communication networks.
The underlying network has a structure that is tunable between a
scale-free growing network with preferential attachments and a
random growing network.  To model realistic situations where
different nodes in a network may have different capabilities, the
message or packet creation and delivering rates at a node are
assumed to depend on the degree of the node.  Noting that
congestions are more likely to take place at the nodes with high
degrees in networks with scale-free character, an efficient
approach of selectively enhancing the message-processing
capability of a small fraction (e.g. $3\%$) of the nodes is shown
to perform just as good as enhancing the capability of all nodes.
The interplay between the creation rate and the delivering rate in
determining non-congested or congested traffic in a network is
studied more numerically and analytically.

\end{abstract}
 \maketitle

\section{Introduction}
Operations in the internet such as browsing webpages in the World
Wide Web (WWW), sending e-mails, transferring files via ftp,
searching for information, and electronic shopping, etc. have
become part of daily life for many people. These activities have
opened up exciting opportunities for sharing information, economic
transformation, and other activities on a global scale
\cite{Rad:1992,Rad:1993}. The internet, however, is not perfect.
For example, intermittent congestion in the internet, similar to
traffic congestion in highway systems, has been observed
\cite{Huber:1997}.  Similar phenomena can be also of relevance in
other communication networks, such as the transportation network
in airlines and the postal service network. A key problem in
communication networks is, therefore, to understand how one can
control congestion and maintain a normal and efficient functioning
of the networks.

Several models of communications in a computer network have been
extensively studied \cite{Li:1989,Leland:1993,Taqqu:1997,
Crove:1997,Falout:1999,Toru:1998,Fuks:1999,Sole:2001,Arena:2001,
Guim:2001,Guim:2002,Woolf:2002,Valv:2002}.  In these models, the
information processors are routers. Their function is to route the
data packets to their destinations.  In a computer network, a node
may be a host or a router. A host can create messages or data
packets for targeted destinations and receive packets from other
hosts. A router finds the shortest path between the origin of the
message and the destination of each packet and forwards the packet
one step closer to the destination along the shortest path in each
time step. The shortest path is the path with the smallest number
of links. Previous studies have mostly been focused on three
different computer network models: (i) the nodes at the edge of
the network are hosts and the inner nodes are routers
\cite{Toru:1998}, (ii) all the nodes are both hosts and routers
\cite{Fuks:1999,Guim:2001,Guim:2002}, and (iii) some of the nodes
are hosts and the rest are routers
\cite{Sole:2001,Woolf:2002,Valv:2002}.  However, these models were
studied with the underlying networks being a two-dimensional
lattice \cite{Toru:1998,Fuks:1999,Sole:2001, Woolf:2002} and a
Cayley tree \cite{Arena:2001,Guim:2001,Guim:2002}.  As the
internet shows a heterogeneous structure with a scale-free degree
distribution \cite{Past:2001,Alex:2002}, a more realistic network
model for communications should be heterogeneous. Besides the
internet and its related networks such as WWW \cite
{Albe:1999,Huber:1999,Kahn:2002} and email networks
\cite{Ebel:2002}, many other networks also show the scale-free
behavior.  These networks include, for example, the telephone
network \cite{Aiello:2000}, the biological network in proteins
\cite{Jeong:2001}, and the networks of sexual contacts
\cite{Fred:2001, Erg:2002}.  Some networks, such as the
collaboration network among scientists \cite{Newm:2001}, show a
mixed feature of scale-free and exponential distributions.  In
fact, the study of the science of complex networks has become an
important interdisciplinary area of research.  The problem of
efficiency in delivering messages or data packets in communication
networks has been addressed recently by Arenas {\em et al.}
\cite{arenas1:2002,arenas2:2004,arenas3:2003}, Moreno {\em et al.}
\cite{moreno1:2003,moreno2:2004,moreno3:2004}, and Zhao {\em et
al.} \cite{zhao:2005}.  Arenas {\em et al.} focused on finding the
optimal network topologies for searches in complex networks, while
Moreno {\em et al.} studied the dependence of the jamming
transitions on routing strategies.  A common feature in previous
studies is that the creation and delivering rates of packages do
not change from node to node.  As the nodes in a complex network
could have very different properties, e.g., degrees, a more
realistic assumption is that the package creation rate and
delivering rate at a node become degree-dependent.  In the
internet, an important site has more users and hence a larger
message or package creation and delivering rates.  A recent study
by Zhao {\em et al.} \cite{zhao:2005} considered the case of
non-uniform package delivering rates, but the creation rate was
taken to be a constant.

In the present work, we study traffic in networks with non-uniform
package creation and delivering rates. An important quantity in
communication networks is the critical package creation rate that
signifies a transition from a non-congested or free flow regime to
a congested regime. Below the critical rate, a non-congested
steady state is reached after the transient in which the data
packets created can be efficiently handled by the nodes. Above the
critical rate, a congested phase is reached where the number of
packets accumulated in the system increases with time.  The value
of critical rate thus measures the capacity of efficient
communication inside the network.  Here, we study the critical
rate in networks where the package creation and delivering rates
are node-dependent.  In particular, we present an efficient
approach to enhance the capacity of communications in scale-free
networks.

The paper is organized as follows. Section II defines the model
with node-dependent package creation and delivering rates. In
Section III, we present results of numerical simulations and study
the interplay between the critical and delivering rates in
determining non-congested or congested traffic in a network.
Section IV explains the observed features in the numerical results
analytically.  We summarize the paper in Sec. V.

\section{Model}

The nodes in a complex network such as the internet may represent
very different entities.  For example, some nodes may just be
individuals and other may represent big companies or universities.
Obviously, different nodes will have different rates of creating
messages.  The nodes, depending on their connectivity to other
nodes and perhaps hardware, also have different rates of
delivering messages.  Here, we present a more realistic model of
communication in complex networks that includes node-dependent
creation and delivering rates.  Our model is a modification on
several previous models
\cite{arenas1:2002,arenas2:2004,arenas3:2003,moreno1:2003,moreno2:2004,moreno3:2004,zhao:2005}.
We assume that for a node $i$ with degree $k_{i}$, the message
creation rate $\lambda k_{i}$ is proportional to its degree, with
$\lambda$ being a constant.  For message delivering, each node
should handle at least one packet or message in each time step.
Therefore, we assume a delivering rate of $1+\beta k_i$ for a node
with degree $k_{i}$, with $\beta \ge 0$ being a parameter of the
model. Our model thus represents the realistic situation that a
busy node with larger $k_{i}$ has higher rates of generating and
delivering messages.

For the underlying network, we use a model in which the exponent
of the degree distribution can be tuned.  A scale-free network
with $P(k) \sim k^{-\gamma}$ can be constructed by incorporating
preferential attachments in a network-growing process
\cite{BA:1999,LL:2002}.  The Barabasi and Albert model
\cite{BA:1999} assumes the probability $\Pi_i$ for a node $i$ to
attract a link from a newly added node to be $\Pi_i \sim k_i$. The
model gives an exponent $\gamma = 3$ \cite{BA:1999} for the degree
distribution.  On the other hand, a random growing network can be
constructed by assuming a node-independent $\Pi_{i}$. Many
networks show characters that are somewhat intermediate of
scale-free and random.  For them, the degree distribution shows a
mixed feature of the two characters \cite{Newm:2001}.  This
implies that the probability $\Pi_i$ of attracting a new link
should contain both preferential and random features.  One of the
present authors proposed a hybrid model \cite{LLYD:2002} in which
$\Pi_i \sim (1-p)k_i+p$, where $0 \le p \le 1$ is a parameter
representing the probability that a newly added node establishes
its new links by random attachments and $(1-p)$ is the probability
that new links are established by preferential attachments.  The
degree distribution was shown to be \cite{LLYD:2002} $P(k) \sim
[k+p/(1-p)]^{-\gamma(p)}$ with an exponent $\gamma(p) = 3 +
p/[m(1-p)]$, where $m$ is the number of new links per node.  The
$p=0$ limit reduces to the $P(k) \sim k^{-3}$ behavior and the $p
\rightarrow 1$ limit gives the random growing network behavior of
$P(k) \sim e^{-k/m}$.  Here, we use this model as our underlying
network for studying communications in networks.

Once the network of a certain value of $p$ is constructed, the
dynamics of creating and delivering messages is implemented as
follows.  Each node plays the dual role of a host and a router,
with its creation and delivering rates assigned according to its
degree.  Details of the dynamics are listed as follows.

\noindent (1) At each time step, a node $i$ has a probability
$\lambda k_i$ of creating a new message or packet with a randomly
chosen destination. If the node has some messages waiting to be
sent, the newly created message will be placed at the end of the
queue.  The queuing messages may be created at some previous time
steps or received from nearby nodes as messages are being sent
along their paths to the destinations.

\noindent (2) Once a packet is created with a chosen destination,
the node (router) will identify the shortest path towards the
destination. If there exist several shortest paths to the
destination, the path is chosen in such a way that the package is
sent to a node that has the instantaneous shortest queue.

\noindent (3) At each time step, a node $i$ has the ability to
forward $(1+\beta k_i)$ packets in the queue at the node on a
first-in-first-out basis to its neighbors which are along the path
to the destinations.  Noting that $\beta k_{i}$ may be an integer
plus a fractional part, the fractional part is implemented as the
probability of delivering additional packets in a time step.

\noindent (4) Messages arriving at a node are queued up for
further delivering.  When a message arrives at its destination, it
is removed from the system.

\noindent The steps are carried out for every node at the same
time. If $\lambda k_i$ is replaced by $\lambda$ and $\beta k_i$ is
replaced by the integral part $int[\beta k_i]$, the above
algorithm will be equivalent to that of Ref. \cite{zhao:2005}.
Here, the fractional part of $\beta k_i$ and $\lambda k_{i}$ are
implemented in a probabilistic way.  Furthermore, if we take
$\beta=0$, the above algorithm will be equivalent to that of Refs.
\cite{Li:1989,Leland:1993,Taqqu:1997,Crove:1997,Falout:1999,
Toru:1998,Fuks:1999,Sole:2001,Arena:2001,Guim:2001,Guim:2002,Woolf:2002,
Valv:2002,arenas1:2002,arenas2:2004,arenas3:2003,moreno1:2003,moreno2:2004,moreno3:2004}.

The parameters $\lambda$ and $\beta$ thus control the number of
messages or packages in the system.  A small (high) value of
$\lambda$ corresponds to fewer (more) packets. For simplicity,
each packet is labeled by two pieces of information: the time of
creation and its destination. Qualitatively, the total number of
packets created at each time step is $\sum_{i=1}^N \lambda k_i
\approx \sum_{i=1}^{N} \lambda \langle k \rangle = 2m\lambda N$
for a growing network of $m$ newly added links per node and a
total of $N$ nodes. At the same time, the maximum number of
packets processed by the nodes is $\sum_{i=1}^N(1+\beta
k_i)\approx (1+2m\beta)N$. When the number of new packets added to
the system equals the number of packages removed upon arrival at
each time step, the network runs in the range of Little law
\cite{Allen:1990} and there is no congestion. For scale-free
networks, the structure is heterogeneous in the sense that some
nodes have many more links.  As messages are sent via the shortest
paths, they are likely to pass through the nodes with more links.
If every node has the same delivering rate, these nodes will have
more accumulated messages and a high chance of jamming. In
contrast, random networks do not have hubs with high degrees and
thus the structure is relatively ``homogeneous".  Therefore,
congestion is easier to occur in scale-free networks than random
networks.  As most of the real-life networks are scale-free, the
control of congestion in these networks is of critical importance.
For example, an intuitive but not so efficient approach is to
increase the value of $\beta$ for the nodes.  We will refer to
this approach of having identical values of $\beta$ for all nodes
as the {\em normal approach}.

The number of nodes in a communication network is typically very
large and there is no central organizer to manage the development
of the whole network.  It is, therefore, very difficult to have a
realistic mechanism to increase $\beta$ for all nodes at the same
time. Realistically, larger companies and academic institutions
could increase their local value of $\beta$ more readily. Noting
that congestion is more likely to occur at the nodes with many
links in scale-free networks, one may significantly reduce
congestion by {\em selectively} increasing the delivering rates of
the nodes with high degrees.  Therefore, we suggest that a
possible cost-effective way to control congestion is to ask the
nodes with larger links to increase their value of $\beta$. Here,
we study a model in which a fraction $f$ of nodes with high
degrees are assigned a finite value of $\beta >0$, and the rest
are assigned $\beta =0$. This models the higher message-processing
capability of the hubs in a network.  We refer to this model as
the {\em efficient approach}. In this model, the maximum number of
packets processed in a time step is $(N+\sum_{i\in f}\beta k_i)$,
where the sum is over the nodes with finite $\beta$. Results of
numerical simulations show that the efficient approach performs
comparably with the normal approach. In the following sections, we
compare results of the two approaches and explain the results
analytically.

\section{Numerical results}

\begin{figure}
\begin{center}
\epsfig{figure=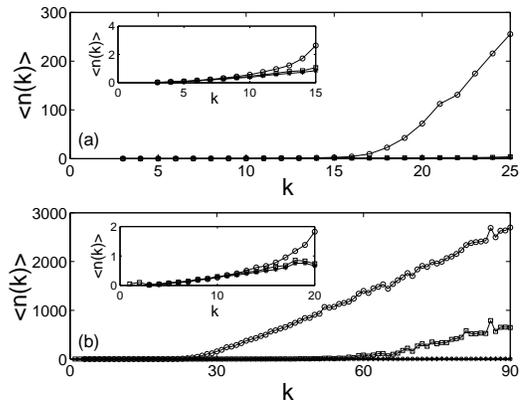,width=0.8\linewidth} \caption{The
average number of packets $\langle n(k) \rangle$ as a function of
the number of links $k$ in networks characterized by $m=3$ and
$N=1000$ for (a) random growing networks ($p=1$) and (b)
scale-free networks ($p=0$). The parameter characterizing the
message creation rate is $\lambda=0.01$.  Results are obtained
after $t=500$ time steps and averaging over $100$ different
realizations for a given set of parameters.  Different symbols
label different packet-processing capabilities: $\beta=0$
(circles), $\beta=0.05$ (squares), and $\beta = 0.1$ (stars).}
\label{fig1}
\end{center}
\end{figure}

The network is constructed as described in Sec. II and in
Ref.\cite{LLYD:2002}, with $N=1000$, $m=3$ and different values of
$p$.  The dynamics of package delivering is then implemented on
the network. We first consider the normal approach in which all
the nodes have the same value of $\beta$. Intuitively, a larger
$\beta$ can assure free traffic flow for a larger creation rate
signified by a larger value of $\lambda$. Here, we fix
$\lambda=0.01$ and take $\beta=0, 0.05$, and $0.1$ to illustrate
the effects.  For $\beta=0$, every node has a creation rate
$\lambda k_i$ depending on $k_{i}$, but the delivering rate of
forwarding at most one message per time step applies to all nodes.
To understand how congestion occurs, we calculate the average
number of packets $\langle n(k) \rangle$ on the nodes with a given
number of links $k$.  This quantity serves to show where are the
longest queues. Numerical simulations show that they are the nodes
with more links, as shown in Fig.\ref{fig1} (circles) for systems
after $t=500$ time steps. The results are obtained by averaging
over $100$ different realizations for a given set of parameters.
Fig.\ref{fig1}(a) shows the results for random growing networks
($p=1$) and Fig.\ref{fig1}(b) shows the results for scale-free
network ($p=0$). It is clear that in both cases, the nodes with
large number of links are more likely to be congested. Comparing
the results in Fig.\ref{fig1}(a) with (b), one sees that the
accumulation of packets in scale-free networks is much pronounced
than that in random networks.  The results indicate that
congestion is much easier to occur in scale-free networks. For
$\beta=0.05$ (squares in Fig.\ref{fig1}), the accumulation of
packets is greatly suppressed in both limits of the underlying
network.  In particular, congestion almost disappeared in the case
of random growing networks. For even higher message-processing
capability $\beta=0.1$ (stars in Fig.\ref{fig1}), congestion
disappeared in both the random and scale-free networks.  These
results also indicate that there exists a critical value
$\beta_{c}$ for a given $\lambda$ so that congestion occurs for
$\beta < \beta_{c}$. We will study the dependence of $\beta_{c}$
on $\lambda$ in networks of different underlying structures
characterized by $p$.

\begin{figure}
\begin{center}
\epsfig{figure=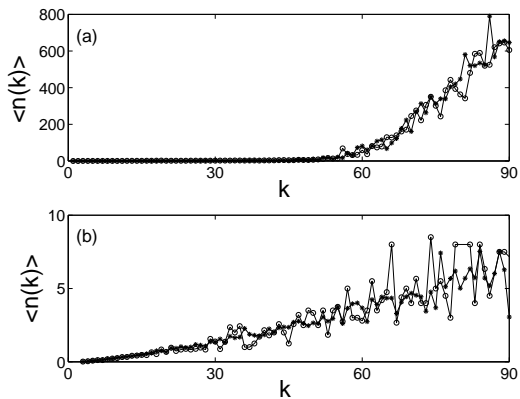,width=0.8\linewidth} \caption{$\langle
n(k) \rangle$ as a function of $k$ for scale-free networks ($p=0$)
with $N=1000$ and $m=3$ for the efficient approach (circles) with
$f=3\%$ and the normal approach (stars).   The parameter
characterizing the message creation rate is $\lambda=0.01$.
Results are obtained after $t=500$ time steps and by averaging
over $100$ different realizations for a given set of parameters.
Two values of $\beta$ are used: (a) $\beta = 0.05$ and (b) $\beta
= 0.1$.} \label{fig2}
\end{center}
\end{figure}

Another point that is worth noticing is that the nodes in a range
of small to intermediate degrees (see inset in Fig.\ref{fig1}) in
the normal approach usually carry fewer messages than the uniform
delivering capacity.  The result implies that it is unnecessary
for these nodes to have a higher delivering capability
characterized by a finite $\beta$.  It leads us to consider the
{\em efficient approach}.  In the scale-free ($p=0$) limit as
shown in Fig.\ref{fig1}(b), the degree distribution is a power
law and the nodes that carry $k \ge 20$ account for only $3\%$ of
all the $N=1000$ nodes.  To illustrate the idea of the efficient
approach, we take $f = 3\%$, i.e., we assign a non-vanishing
$\beta$ only to nodes that have $k \ge 20$.  Figure \ref{fig2}
compares results of the normal approach (stars) and the efficient
approach (circles) for two values of $\beta$.  Obviously the
difference in $\langle n(k) \rangle$ between the two approaches
is small in scale-free networks.  As the efficient approach does
not require all nodes to be equipped with the same capability, it
represents a more practical and cost-effective way to avoid
jamming.  In contrast, we note that the efficient approach does
not work so well in random networks ($p=1$).  It is because the
degree distribution is narrower compared with the $p=0$ case.
Therefore, the queues are more evenly distributed among the nodes
and congestion is not restricted to the nodes among the highest
degrees.  It is thus necessary to assign a finite $f$ to a larger
fraction of nodes to avoid congestion, and the two approaches
become similar.

\begin{figure}
\begin{center}
\epsfig{figure=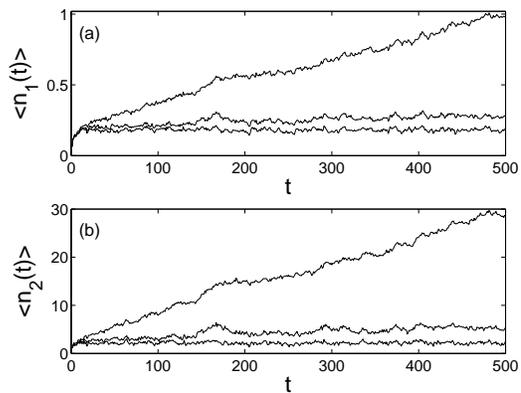,width=0.8\linewidth} \caption{(a) The
average number of messages per node and (b) the average number of
messages among the top $3\%$ nodes with the highest degrees as a
function of time in a scale-free network ($p=0$) for three
different values of $\beta$.  The lines from top to bottom refer
to $\beta = 0.05$, $0.059$, and $0.07$, respectively. The other
parameters are $\lambda=0.01$, $f=3\%$, $N=1000$, and $m=3$.}
\label{fig3}
\end{center}
\end{figure}

We define $\langle n_{1}(t) \rangle$ to be the average number of
messages per node.  In the congested regime, $\langle n_{1}(t)
\rangle$ increases with time $t$.  In the non-congested regime,
$\langle n_{1}(t) \rangle$ fluctuates around a constant. The
slope of $<n_1(t)>$ after the transient can thus be used to
determine $\beta_{c}$
\cite{arenas3:2003,moreno2:2004}. For a given $\lambda$, the
slope gradually decreases as $\beta$ increases.  The value of
$\beta$ that the slope becomes zero gives $\beta_{c}$. For
$\beta>\beta_c$, the slope remains zero.  Figure \ref{fig3}(a)
shows typical results with $\lambda = 0.01$ for three different
values of $\beta$ within the efficient approach ($f = 3\%$).  For
$\beta = 0.05$, $\langle n_{1}(t) \rangle$ increases with time
without bound.  The critical value is found to be $\beta_{c} =
0.059$ where the slope vanishes.  For $\beta = 0.7 > \beta_{c}$,
the slope remains zero.  As congestion mainly occurs at the nodes
with large degrees, the number of messages $\langle n_{2}(t)
\rangle$ averaged over the $3\%$ of nodes should also show a
similar behavior with time.  It is indeed the case (see
Fig.\ref{fig3}(b)).

\begin{figure}
\begin{center}
\epsfig{figure=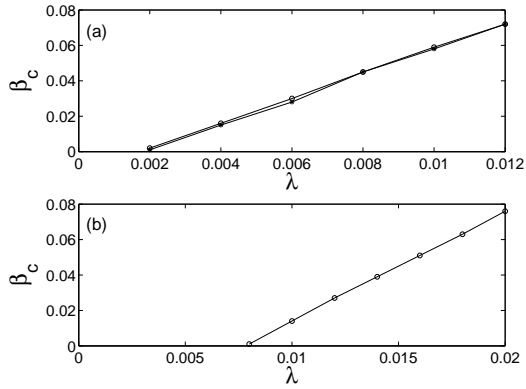,width=0.8\linewidth} \caption{$\beta_c$
as a function of $\lambda$ for (a) scale-free networks ($p=0$) of
$N=1000$ and $m=3$ within the normal (circle) and efficient
(stars) approaches ($f=3\%$); and (b) random growing networks
($p=1$).} \label{fig4}
\end{center}
\end{figure}

Next, we study the dependence of $\beta_c$ on the creation rate
characterized by $\lambda$ in scale-free and random growing
networks.  Fig.\ref{fig4}(a) shows the results in the scale-free
limit ($p=0$) for both the normal (circles) and efficient (stars)
approaches. For small $\lambda$, $\beta_{c}$ vanishes as the
default delivering rate of one message per time step is already
sufficient to handle the small message creation rate.  For the
range of $\lambda$ shown in the figure, $\beta_{c}$ increases
linearly with $\lambda$ and the two approaches give similar
results.  This again shows that the efficient approach performs as
good as adjusting $\beta$ across the whole network. It should be
noted that for larger values of $\lambda$ (beyond the range shown
here), assigning a finite $\beta$ to only the top $3\%$ of nodes
may not be sufficient to avoid congestion.   For random growing
networks ($p=1$), we show $\beta_{c}(\lambda)$ in
Fig.\ref{fig4}(b) only for the normal approach, as the efficient
approach becomes similar to the normal approach.  Qualitatively,
$\beta_{c}(\lambda)$ shows a similar behavior to that in
scale-free networks.  Quantitatively, $\beta_{c}=0$ for a larger
range of $\lambda$ in random networks and the slope of the linear
dependence in $\beta_{c}(\lambda)$ is smaller.  It is because the
nodes in random networks are more ``homogeneous" and a queue will
not emerge at the hubs for small $\lambda$ as in the case of
scale-free networks.  The function $\beta_{c}(\lambda)$ in
Fig.\ref{fig4}(a) also divides the $\beta$-$\lambda$ space into
two regions.  The region above the line represents a non-congested
or free flow regime and that below the line represents a congested
regime. Thus for given $\lambda$, one can go from a congested to a
non-congested regime by increasing $\beta$.  Similarly, for a
given $\beta$, one can go from a non-congested regime to a
congested regime by increasing $\lambda$.  Although we only
present results for networks with $N=1000$ nodes, we have checked
that the linear dependence of $\beta_{c}$ on $\lambda$ also holds
for networks with larger $N$.

\section{Theoretical explanation}

In this section, we aim at explaining the behavior of
$\beta_{c}(\lambda)$ quantitatively.  As discussed, $\langle
n_{1}(t) \rangle$ behaves differently for $\beta < \beta_{c}$ and
$\beta \ge \beta_{c}$.  In the non-congested or free flow regime,
the steady state satisfies the Little's law \cite{Allen:1990},
which states that the number of delivered messages is balanced by
the number of newly created messages.  This suggests a way to
estimate $\beta_{c}(\lambda)$ for a given underlying network
characterized by the parameter $p$.

Consider the node with the highest degree $k_{max}(p)$ where
messages are most likely to accumulate. At the critical value
$\beta_{c}$, the node can handle $1+\beta_c k_{max}(p)$ packets
per time step, while the number is smaller for other nodes. For
scale-free networks ($p=0$), $k_{max}\approx m N^{1/(\gamma-1)}$
\cite{AB:2002}.  We note that the packets at a node $i$ originate
from two different sources: those created at node $i$ and those
passing by node $i$. The creation rate $\lambda k_i$ is linear in
$k_{i}$.  The packets passing by are more likely to go through the
nodes with higher degrees, and hence the number of packets passing
by a node will be some nonlinear function of its degree. With
these considerations, we approximate the average number of packets
at the critical value $\beta_{c}$ at some node $i$ as
$\alpha(k_i,p)(1+\beta_ck_{max}(p))k_i/k_{max}(p)$, where $0<
\alpha(k_i,p)\le 1$ and $\alpha(k_{max},p)=1$ is a nonlinear
decreasing function of $k_i$ that reflects the contribution of
messages passing by the node.  For the case of $p=0$, noting that
there are only $1+\beta_c k_{max}(0)$ packets at the nodes with
$k_{max}$, the average number of packets at the nodes with small
and intermediate degrees will be less than one.  This implies that
there is not enough packets for the parameter $\beta$ to take
effect at these nodes.  Therefore, we expect the expression
$\alpha(k_i,p)(1+\beta_ck_{max}(p))k_i/k_{max}(p)$ to be a good
approximation for both the normal and efficient approaches in
scale-free networks.  On the other hand, there are $2 m \lambda N$
newly created packets in each time step.  Let $h(p)$ be the
diameter of the network which measures the average number of nodes
that a packet passes through on its way to its destination,
including the destination itself.  If the system is in the
non-congested regime, there are a total of $h(p)2m\lambda N$
messages in the system. To avoid a queue at any node and hence
congestion, all the messages should be handled by the nodes in a
time step.  Thus, we have
\begin{equation} \label{eq:Little}
h(p)2m\lambda
N=\sum_{i=1}^N\alpha(k_i,p)(1+\beta_ck_{max}(p))k_i/k_{max}(p).
\end{equation}
Writing
$\sum_{i=1}^N\alpha(k_i,p)(1+\beta_ck_{max}(p))k_i/k_{max}(p)\equiv
\alpha_1(p)\sum_{i=1}^N(1+\beta_ck_{max}(p))k_i/k_{max}(p)$, we
then have
\begin{equation} \label{eq:Critical_0}
\beta_c(\lambda)
=\frac{h(p)\lambda}{\alpha_1(p)}-\frac{1}{k_{max}(p)}.
\end{equation}
From Eq.(\ref{eq:Critical_0}), it follows that (i) $\beta_{c} > 0$
only when $\lambda$ is sufficiently large, (ii) for a given
structure of the network (fixed $p$), $\beta_{c}$ increases
linearly with $\lambda$, and (iii) the slope of
$\beta_{c}(\lambda)$ depends on the underlying network structure
characterized by $p$.  All these features agree with those
observed in the numerical results (see Fig.\ref{fig4}).

\begin{figure}
\begin{center}
\epsfig{figure=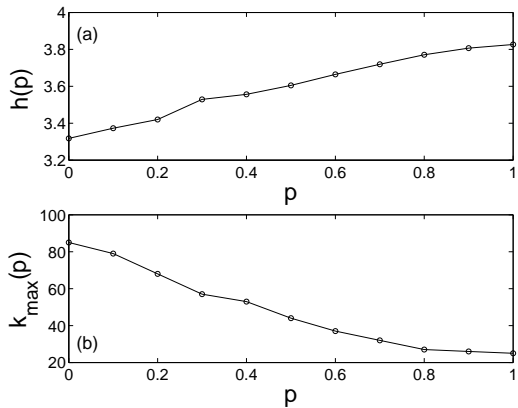,width=0.8\linewidth} \caption{(a) The
diameter $h(p)$ and (b) the highest degree $k_{max}(p)$ in a
network as a function of the underlying network structure
characterized by the parameter $p$.  The networks have $N=1000$
nodes and $m=3$.} \label{fig5}
\end{center}
\end{figure}

Equation (\ref{eq:Critical_0}) can be applied to estimate
$\beta_{c}$, if we know $h(p)$, $k_{max}(p)$, and $\alpha_1(p)$.
The diameter $h(p)$ can be calculated using the method in Ref.
\cite{LL:2002}. Figure \ref{fig5}(a) shows $h(p)$ over the whole
range of $p$.  It increases only slightly as $p$ increases.  On
the other hand, $k_{max}(p)$ drops sensitively with $p$ as shown
in Fig. \ref{fig5}(b).  Since $\alpha_1(p)$ depends only on $p$,
it can be determined by using numerical results of $\beta_{c}$ for
given $\lambda$.  For example, $\beta_{c}(\lambda=0.01) = 0.059$
in the scale-free ($p=0$) limit.  Together with $h(0)=3.32$ and
$k_{max}(0)=85$ (see Fig.\ref{fig5}), Eq.(\ref{eq:Critical_0})
gives $\alpha_1(0)\approx 0.4522$ and the slope of the line
$\beta_{c}(\lambda)$ is $h(p)/\alpha_1(p) = 7.34$.  Similarly,
$\beta_{c}(\lambda = 0.012) = 0.027$ in the random network limit
($p=1$).  Together with $h(1)=3.82$ and $k_{max}(1)=25$ (see
Fig.\ref{fig5}), Eq.(\ref{eq:Critical_0}) gives
$\alpha_1(1)\approx 0.6842$ and the slope of the line
$\beta_{c}(\lambda)$ to be $5.58$.  These values are in reasonable
agreement with the slopes in the plots in Fig.\ref{fig4}. Equation
(\ref{eq:Critical_0}) also shows that $\beta_{c} = 0$ for $\lambda
< \lambda_{min} = \alpha_1(p)/[h(p)k_{max}(p)]$.  Using the
extracted values of the parameters, we get $\lambda_{min} =
0.0016$ for $p=0$ and $\lambda_{min} = 0.0072$ for $p=1$. These
values are consistent with the results in Fig.\ref{fig4}.

\section{Conclusions}
In network communications, a simple way to control network traffic
is to limit the length of the queues \cite{Csa:1994}, e.g. by
source quenching, random dropping, fair queueing, etc.  This will,
however, increase the average delivering time.  As many real-life
networks are heterogeneous networks and many shortest paths
between any two nodes pass through the nodes with high degrees, it
will be these nodes that control the network traffic.  Thus we
study the strategy of enhancing the message delivering capability
selectively at the nodes with high degrees.  We found that the
strategy works well in networks with scale-free character and it
is a highly cost-effective way to avoid network congestion. This
idea is in line with the recent results in
Ref.\cite{Adilson:2004,Zoltan:2004}.

The major difference in network congestion in a scale-free network
and a random growing network is that the scale-free network has
hubs, i.e., nodes that are connected to many other nodes.  The
degree distribution in a scale-free network follows a power law
for large networks.  In a random network, the degree distribution
is relatively narrower and the degrees of the nodes do not differ
by much.  For identical message creation rate and delivering rate
at the nodes, it is then expected that congestion will take place
mostly at the nodes of high degrees in a scale-free network.  For
a random network, congestion may take place at more places across
the network.  Strategically enhancing the message-processing
capability at the high-degree nodes in a scale-free network as in
the efficient approach studied in the present work will greatly
enhance network traffic.  This strategy also makes good use of the
power-law degree distribution in that it is sufficient to allocate
resources to enhance the capability of a small fraction of nodes
with high degrees in a network in order to avoid traffic
congestion.  If we carry out the same strategy to a random growing
network, a much larger fraction of nodes will be involved and
hence the cost-effectiveness will be lowered.

In summary, we have constructed and studied a model of
communications in complex networks.  We use a network model that
can be tuned from the scale-free preferential growing network
limit to the random growing network limit.  Our model assumes a
message creation rate $\lambda k_i$ that depends on the degree of
a node.  Each node also has a message delivering rates of $1+\beta
k_i$. The model thus represents a step towards a more realistic
modelling of traffic congestion in communication networks in that
it incorporates the different capacities of the nodes in creating
and handling messages. In particular, we studied an efficient
approach that increases the communication capacity in scale-free
networks.  Numerical results indicate that our efficient approach
of selectively enhancing the delivering rate in a small fraction
of nodes performs as good as enhancing the capability of all the
nodes in the network.  Considering the cost of enhancing the
delivering rate at a node, the present scheme will be highly
cost-effective.  We also studied the dependence of the critical
value of $\beta$, which characterizes the message delivering rate,
on the parameter characterizing the message creation rate
$\lambda$.  The function $\beta_{c}(\lambda)$ divides the
$\beta$-$\lambda$ space into two regions of physically different
characters: non-congested or free flow regime and congested
regime.  Analytically, we derive an expression of
$\beta_{c}(\lambda)$ based on the idea that all the messages in
the system should be handled by the nodes in the non-congested
regime.  The analytic expression captures all the features
observed in numerical results.

\acknowledgments This work was supported in part by the National
Science Foundation of China under Grant No. 10475027 (Z.L.), SRF
for ROCS, SEM under Grant No. 44020460 (Z.L.), and the Research
Grants Council of the Hong Kong SAR Government under Grant No.
CUHK-401005 (P.M.H.). This work was partially completed during a
visit of Z.L. to CUHK which was supported by a Direct Grant of
Research from CUHK.

\end{document}